# Hybrid integrated optical waveguides in glass for enhanced visible photoluminescence of nanoemitters

Josslyn Beltran Madrigal,[1,] Ricardo Tellez-Limon,[1,2,] Florent Gardillou,[3] Denis Barbier,[3]
Wei Geng,[1] Christophe Couteau,[1,4] Rafael Salas-Montiel,[1] and Sylvain Blaize[1,*]

[1]Laboratoire de Nanotechnologie et d'Instrumentation Optique (LNIO), ICD CNRS UMR6281, Université de Technologie de Troyes, 12 Rue Marie Curie CS 42060, 10004 Troyes, France
[2]Current address: CONACYT - Unidad Monterrey, Centro de Investigación Científica y de Educación Superior de Ensenada, Alianza Centro 504 , Apodaca, NL 66629 ,Mexico
[3]TEEM Photonics, 61 Chemin de Vieux Chêne, 38240 Meylan, France
[4]CINTRA CNRS-NTU-Thales UMI 3288 & Electrical and Electronic Engineering School, Nanyang Technological University, Singapore 639798, Singapore
*Corresponding author: sylvain.blaize@utt.fr

ABSTRACT

Integrated optical devices able to control light−matter interactions on the nanoscale have attracted the attention of the scientific community in recent years. However, most of these devices are based on silicon waveguides, limiting their use for telecommunication wavelengths. In this contribution, we propose an integrated device that operates with light in the visible spectrum. The proposed device is a hybrid structure consisting of a high-refractive-index layer placed on top of an ion-exchanged glass waveguide. We demonstrate that this hybrid structure serves as an efficient light coupler for the excitation of nanoemitters. The numerical and experimental results show that the device can enhance the electromagnetic field confinement up to 11 times, allowing a higher photoluminescence signal from nanocrystals placed on its surface. The designed device opens new perspectives in the generation of new optical devices suitable for quantum information or for optical sensing.

INTRODUCTION

From telecommunication systems that implement kilometerlong optical fibers, to micrometric and nanometric systems for the detection of biochemical substances, integrated optical devices play an essential role for the control of light−matter interactions. Because of their large number of applications, these devices have been widely studied over the last decades. Examples of these applications are multiplexers [1–4], optical switches [5], isolators [6,7], modulators [8,9], interferometers [10–12], and gas detectors [13–15], to name a few. Based on new fabrication techniques, an important aim is the miniaturization of these optical circuits in order to confine light in subwavelength volumes.

In recent years, several integrated optical devices have been successfully used to excite light sources with dimensions of a few nanometers. Some examples are the excitation of nanoemitters by surface plasmon polaritons (SPPs) [16], localized surface plasmons in nanoantennas [17–19], optical cavities [20,21], nanocrystals, [22,23] nanowires [24–26], and metallic waveguides [27–29]. However, despite the fact that the absorption/emission wavelengths of the nanoemitters used in these works rely on the visible spectrum (CdSe or CdS quantum dots and nitrogen vacancy centers in diamond), the materials used for their excitation have high losses in this spectral range (silicon and noble metals). Thus, the challenge to be faced is to design integrated devices for strong light confinement using low-loss materials.

For this purpose, a possible platform for integrated optics would be the so-called ion-exchanged glass waveguides (IEWs) [30,31]. These waveguides present many interesting advantages, such as low loss and low birefringence, low fabrication cost, compatibility with standard optical fibers, and high stability [32–34]. These photonic structures have been used to develop passive devices like interferometers or multiplexers [35,36] as well as active devices, such as lasing amplifying media [37] and lasers [38,39] based on Erbium ion doping.

Nevertheless, IEWs have low index contrast between the core and the cladding, leading to low light confinement. Because of this reason, these waveguides are not commonly used to excite nanostructures, and only a few examples can be found in literature, like the excitation of ammonia molecules [40]. So far, however, IEWs have not been used in combination with fluorescent molecules or nanocrystals. In this contribution, we demonstrate that IEWs in combination with a well-chosen high-index dielectric thin film acts as an efficient light−matter coupler for the excitation of nanoemitters placed atop. The theoretical design of this hybrid structure was done with the Fourier modal method (FMM), and the fabricated sample was experimentally characterized with the near-field scanning optical microscopy (NSOM) technique.

To prove the performance of our integrated device, we placed on its surface cadmium selenide/cadmium sulfide (CdSe/CdS) nanocrystals and measured their photoluminescence (PL) and fluorescence lifetime. We theoretically and experimentally demonstrated an enhancement of one order of magnitude in the PL emission of the NCs placed on top of the hybrid structure, relative to the PL emission of the same NCs placed only on top of the IEW.

## DESCRIPTION OF THE DESIGNED STRUCTURE

The hybrid structure (HS) consists of a titanium dioxide (TiO$_2$) slab waveguide (n ~2 at λ 540 nm) placed on top of an ion-exchanged waveguide in glass manufactured by the company Teem Photonics [41] [Fig. 1(a)]. The IEWs were fabricated by the substitution of sodium ions for silver ions in a glass substrate that locally increases the index of refraction of the glass [30,42]. A photolithography process was used to define diffusion windows of 1 μm. As a result of the thermal diffusion of the sodium/silver ions, the core of the waveguides presents a gradient-index distribution [Fig. 1(b)]. The high refractive index layer of TiO$_2$ of thickness 85 nm was thermally evaporated in a region patterned by electron-beam lithography [Fig. 1(c)].

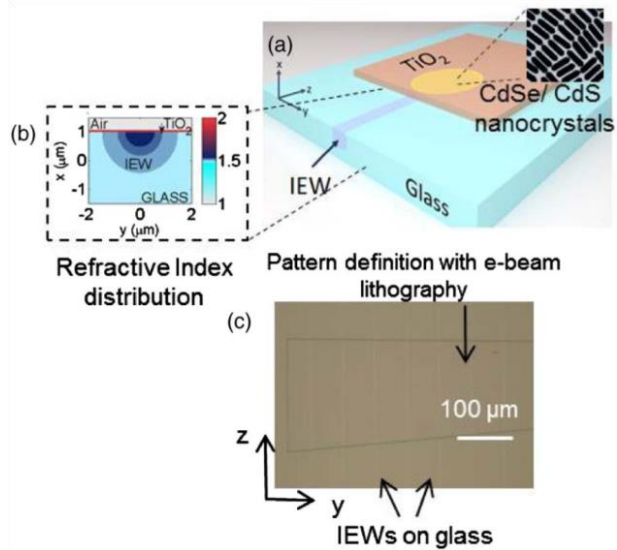

Fig. 1. (a) Schematic representation of the hybrid structure consisting of a TiO$_2$ layer placed on top of an IEW with a nanocrystals solution on its surface. The inset is a schematic representation of the CdSe/CdS nanocrystals. (b) Refractive index distribution of the hybrid structure. (c) Top view of the fabricated layer atop the glass substrate containing several IEWs.

The highly fluorescent dot-in-rod CdSe/CdS colloidal nanocrystals (NCs) were synthesized in liquid phase following a seeded growth approach. They consist of a CdSe spherical core of 2.7 nm in diameter, surrounded by an elongated shell of CdS to form a rod shape with dimensions of 50 nm in length and 7 nm in diameter. A high concentration of NCs ($10^{-6}$ mol/L) was mixed in a poly (methyl methacrylate) (PMMA) solution and randomly distributed on top of the hybrid structure. The top view of the HS imaged with an optical microscope is shown in Fig. 1(c). The IEWs parallel to the z axis are separated by a distance of 51 μm in the y direction, and the area delimited with dark lines corresponds to the TiO$_2$ layer.

## RESULTS

Figure 2(a) illustrates the evolution of a hybrid mode as it propagates along the HS formed by the IEW and the TiO$_2$ slab. Due to the proximity of both structures, the evanescent wave of the propagating mode of the IEW is able to excite the modes of the TiO$_2$ slab, resulting in a two waveguide coupled system. This interaction gives rise to an hybrid mode characterized by a periodic energy transfer between both waveguides [43,44] with a coupling length $L_c$. We then define the enhancement factor (EF) of the electric field as $EF = |E(x_1; z_1; \omega)|^2 / |E_0(x_0; z_0; \omega)|^2$, where $E(x_1; z_1; \omega)$ is the electric field at the point of maximum intensity of the hybrid mode at 5 nm from the surface of the $TiO_2$ layer, and $E_0(x_0; z_0; \omega)$ the electric field at a reference point 5 nm above the IEW, as indicated in Fig. 2(a).

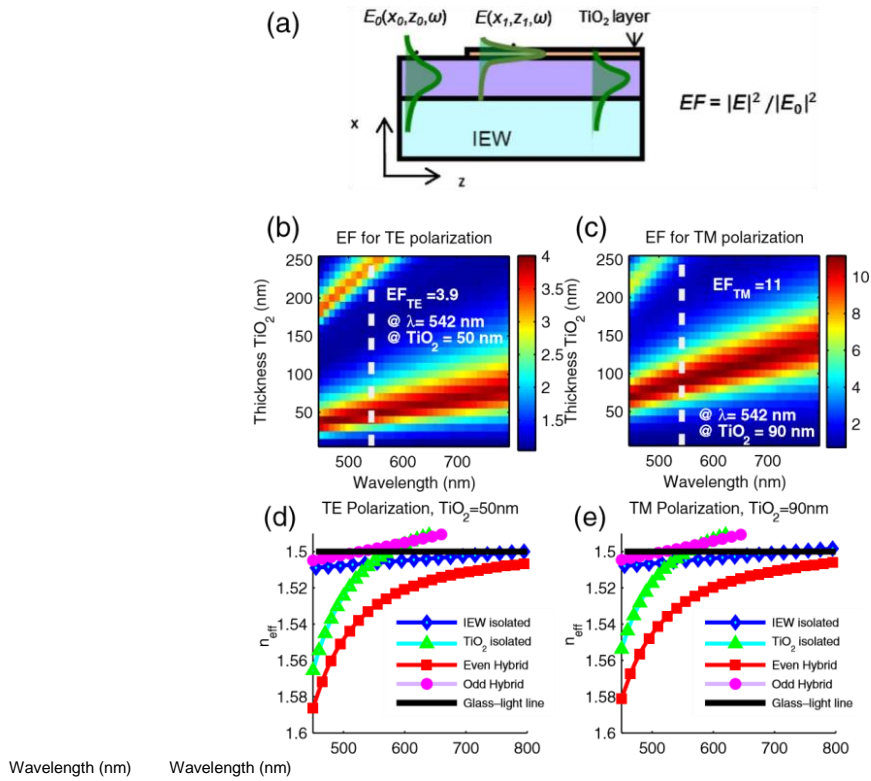

Fig. 2. (a) Schematic of the cross section of the structure and definition of the EF. Calculated EF as a function of the wavelength and thickness of the $TiO_2$ layer for the (b) TE and (c) TM incident ion-exchanged optical modes. The white guidelines and corresponding tags indicate the maximum EF factor and thickness of the $TiO_2$ layer at the operation wavelength of $\lambda=542$ nm. Dispersion relation of the isolated ion-exchanged waveguide on glass (blue diamonds), of an isolated $TiO_2$ slab waveguide on glass (green triangles), and of the coupled hybrid structure red rectangles and pink circles) for (d) the TE mode and (e) the TM mode. The dispersion of the hybrid structure presents an anticrossing point at 542 nm.

Figures 2(b) and 2(c) show the EF as a function of the wavelength and thickness of the $TiO_2$ layer for incident transverse electric (TE) and transverse magnetic (TM) fundamental modes, respectively. For the TE mode, the electric field is oriented along the y axis and for TM along the x axis. These values were numerically calculated by modeling the beam propagation with the Fourier modal method (see Appendix A).

For a wavelength $\lambda = 542$ nm, the maximum enhancement factor obtained for TE polarization is $EF_{TE} = 3.9$ corresponding to a $TiO_2$ layer of thickness 50 nm, while for TM polarization, $EF_{TM} = 11$ for a thickness of 90 nm [Figs. 2(b) and 2(c)]. As the number of modes propagating in the $TiO_2$ layer increases with its thickness, a second maximum appears for thicker layers but with smaller EF since the energy of the electromagnetic field is distributed in both modes. Due to the invariant geometry of the system along the y direction and the orientation of the electric field of both polarizations, the EF is smaller for TE polarization because the system is less sensitive to material perturbations in the vertical direction (x axis). These results are advantageous over a plasmonic structure. For instance, by replacing the $TiO_2$ layer with a gold layer, the maximum enhancement factor was $EF_{SPP} = 5.87$, and it occurs for a layer of thickness of 10 nm at a wavelength around $\lambda = 800$ nm for TM polarization (due to the geometry of the system, it is not possible to excite the SPP resonance with a TE mode). Besides, the propagation distance for the metallic layer was about 25 μm, while for $TiO_2$ it was longer than 200 μm.

Figures 2(d) and 2(e) represent the dispersion curves for TE and TM polarizations. The fundamental mode of the IEW is represented by blue diamonds, and the mode supported by the $TiO_2$ layer on top of a glass substrate (without the presence of the IEW) is represented by green triangles. The red squares and pink circles correspond to the even and odd branches of the hybrid mode, respectively. This relation represents the effective indices of the eigenmodes as a function of the wavelength. The crossing phenomenon at λ =542 nm represents the phase matching between the modes of the IEW and the $TiO_2$ layer, allowing a maximum energy transfer between them (where EF reached its maximum value).

Figures 3(a) and 3(b) show the propagation of light through the HS when the HS is excited with the fundamental TE and TM modes of the IEW, respectively. These results were obtained with the FMM (see Appendix A) for a $TiO_2$ layer with a thickness of 85 nm (close to the thickness of the fabricated HS). From the numerical results, EF =2.4 for the TE mode and EF =10.2 for the TM mode. In addition, the coupling length ($L_c$) from IEW to the $TiO_2$ layer is $L_c$ =2.1 µm for the TE mode and $L_c$ =6.9 µm for the TM mode.

The intensity of the electromagnetic field confined to the surface of the fabricated HS (top view) was measured with the NSOM technique (see Appendix A) and their maps are shown in Figs. 3(c) and 3(d) for the incident TE and TM modes at λ =542 nm, respectively. From these maps, we measured EF =1.9 and $L_c$ =2 µm for TE polarization, while for TM polarization EF =10.2 and $L_c$ =7.1 µm. The calculated and measured EF agrees [Figs. 3(e) and 3(f)].

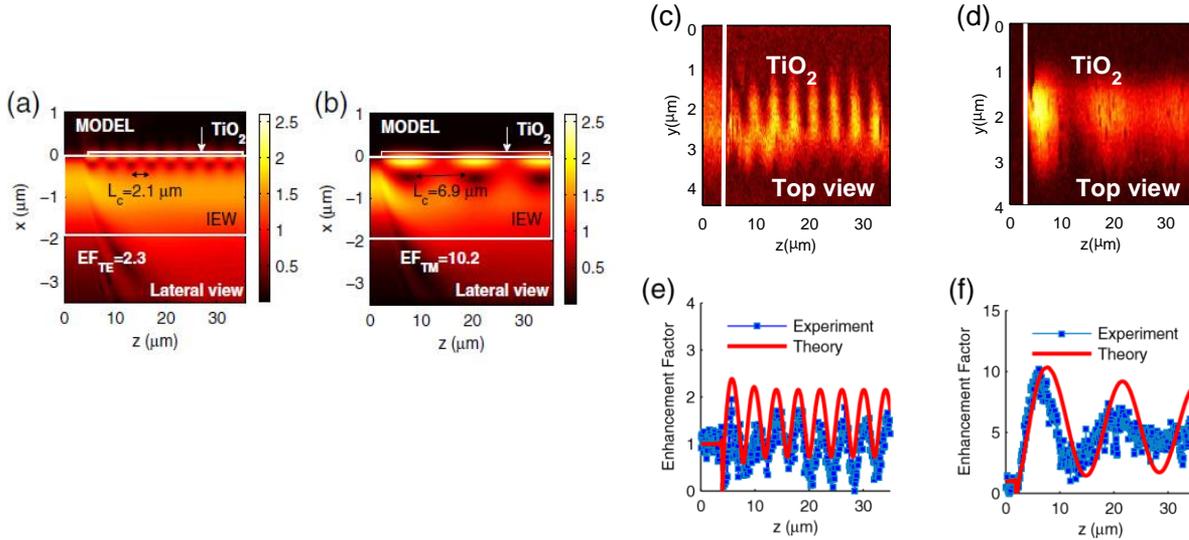

Fig. 3. Comparison between experimental and numerical results for incident TE (left column) and TM (right column) modes propagating through the hybrid structure at a wavelength of 542 nm. The thickness of the $TiO_2$ layer is 85 nm. (a),(b) Near-field maps of the lateral cross section computed with the FMM. (c),(d) Near-field intensity maps measured with the NSOM technique on the surface of the sample. (e),(f) Comparison of the numerical and experimental enhancement factors on top of the HS.

After the optical near-field characterization of the HS, we performed PL measurements of CdSe/CdS nanocrystals placed on top of the integrated device (see Appendix A). The results are presented in Figs. 4(a) and 4(b), corresponding to an excitation with TE and TM modes, respectively, from the input of the IEW at a wavelength λ =532 nm. In these images, the excitation pump was spectrally filtered out, such that the CCD images only provide information of the PL of the NCs. The nanocrystals mainly emit at λ =590 nm, as can be seen by the spectra of Figs. 4(c) and 4(d). No blinking effects were observed due to the high concentration of the NCs. We define the enhancement factor of the PL of the nanocrystals ($EF_{PL}$) as the ratio between the PL of the nanocrystals placed on top of the $TiO_2$ layer and the PL of the NCs directly placed on top of the IEW. This enhancement in the PL ($EF_{PL}$) is 1.3±0.2 times for the TE mode and 7.3±0.1 times for the TM mode. In the HS region, we measured a coupling length of $L_c$ ≃1.2 µm for the TE mode and $L_c$ ≃2.8 µm for TM.

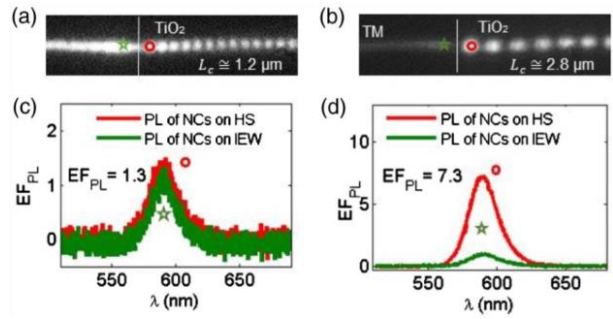

Fig. 4. Far-field observation of light emission from CdSe/CdS nanocrystals placed on top of the HS for (a) TE and (b) TM polarizations when exciting the structure from the inlet of the IEW. The PL spectra were extracted from the image at the surface of the IEW (star point) and at the surface of the hybrid section of the structure (circle point) for (c) the TE-polarized mode and (d) the TM-polarized mode. A PL enhancement factor of 1.3±0.2 times is obtained for TE polarization and 7.3±0.1 times for TM polarization at the wavelength of maximum emission of the nanocrystals (i.e., at 590 nm).

These results differ from the values of Fig. 3 not only because of the different operation wavelength but mainly because of the effective thickness of the high-index guiding layer. Indeed, the solution of NCs and PMMA deposited on the surface of the HS had a thickness of ~30 nm (AFM measurement). This value corresponds to half of the length of the rods due to their random positioning. Furthermore, at this wavelength, the average refractive index of this solution is around 2 ($n_{PMMA}$ =1.5 and $n_{NC}$ =2.5), which is close to index of refraction of the $TiO_2$. Hence, when numerically simulating the beam propagation of the HS for a layer of $TiO_2$ with a thickness of 120 nm, we obtained EF =1.4 and $L_c$ =1.1 μm for the TE mode and EF =6.6 and $L_c$ =3 μm for TM. These values agree with the experimental observations.

Figure 5 shows the fluorescence signal of the CdSe/CdSe nanocrystals when they were excited at normal incidence with a pulsed laser centered at λ 405 nm focused on the sample with a confocal system (see Appendix A). The NCs were excited at four strategic points on the sample: NCs placed on glass, on IEW, on $TiO_2$-on-glass, and on the HS. The highest fluorescence intensity was obtained for nanocrystals placed on top of the HS with a signal six times higher than the one measured with the NCs placed on top of the glass substrate.

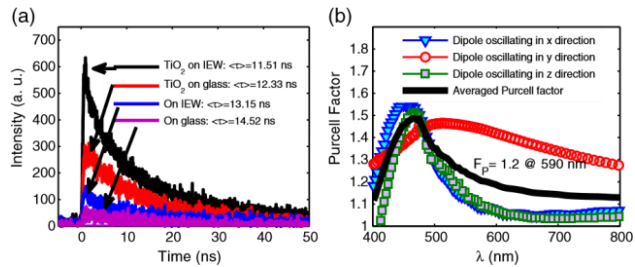

Fig. 5. (a) Lifetime measurements at λ=590 nm of CdSe/CdS nanocrystals on glass, on WG in glass, on $TiO_2$ on glass, and on HS. The NCs were excited at λ=405nm. (b) Purcell factor obtained with the FDTD method for a dipole oriented along the three coordinate axes (blue, red, and green) and averaged Purcell factor (black).

The measurements of the fluorescence lifetime at the emission wavelength of the NCs (λ =590 nm) are presented in Fig. 5(a). The line shape of the decay intensity was approximated to a single exponential $I=I_0 e^{-t/\tau}$ with τ as the NC's lifetime [45,46]. According to the analysis in reference [47], the Purcell factor ($F_P$) is defined in terms of the lifetime and local density of excited states (LDOS) of the nanocrystals as $F_p = \tau_0/\tau = LDOS/LDOS_0$, where τ is the modified lifetime (with the HS), and $\tau_0$ is the lifetime for the NCs on top of the glass substrate.

For nanocrystals placed on top of the HS, we measured a Purcell factor of $F_p$ =1.3. According to our numerical results (see Appendix A), by averaging the Purcell factor of a single NC oriented in the three coordinate axes and normalized to one placed on top of a glass substrate, we found $F_p$ =1.2 [Fig. 5(b)]. A well-known fact is that a proper orientation of the NC will lead to a higher Purcell factor, as depicted in Fig. 5(b), which is a challenging situation requiring further engineering.

## CONCLUSION

We demonstrated the improvement of light confinement in a common ion-exchanged glass waveguide through the implementation of a high-index layer of $TiO_2$ placed on its surface. By making use of the Fourier modal method, we demonstrated that the hybrid structure can provide an enhancement of the electromagnetic field of up to 11 times for TM polarization and 3.9 for TE at λ =542 nm. The theoretical results were experimentally validated by the optical characterization of the hybrid structure with the NSOM technique.

We also showed the performance of our device as an integrated waveguide platform for the excitation of CdSe/CdS nanocrystals. The HS allowed us to increase the photoluminescence of the nanocrystals up to 7.3 times for TM polarization. Additionally, by measuring the fluorescence lifetime of the NCs, we observed an increase of 30% in the Purcell factor.

Our structure is a proof-of-concept that opens new perspectives in the design of integrated nanophotonic devices requiring of the near-field enhancement of light for the excitation of nanostructures that can lead to outstanding applications, such as quantum information, telecommunications, light generation, imaging, data storage, sensing, and biomedical applications.

## APPENDIX A: METHODS

### 1. Fourier Modal Method

The Fourier modal method is a rigorous method that solves the Maxwell equations in the frequency domain. It is based on the solution of an eigenvalue problem arising from the expansion in the Fourier series of the electromagnetic field as well as the material properties of a unit cell of a periodic structure. A detailed explanation of the method can be found in Refs. [48–50].

To solve a nonperiodic multilayered system (beam propagation), perfectly matched layers were included in both sides of the unitary cell to avoid reflections into the region of interest. The propagation constants and effective indices of the modes (dispersion curves) supported by the structure were directly obtained from the solution of the eigenvalue problem [51].

To numerically model the gradient-index profile of the ion-exchanged waveguide within the framework of the Fourier modal method, we used a multilayered medium consisting of five layers. Each layer with thickness and refractive index matching the index profile is described in Fig. 2(a). Each layer with a refractive index following the index profile is described in Fig. 1(b). The refractive index of the $TiO_2$ layer placed on top of the glass substrate (without IEW) was experimentally measured by ellipsometry in our facilities.

### 2. Near-Field Scanning Optical Microscopy

The near-field measurements were done through near-field scanning optical microscopy in an apertureless configuration [52,53] by coupling light from a He–Ne laser ($\lambda$ 542 nm) to an optical fiber with a pigtailed polarization controller stage. We used a single-mode polarization-maintained microlensed fiber to couple light into the IEW. In order to measure the transmitted power, the outlet face of the IEW was coupled to a second optical fiber with the same characteristics as the input fiber, which was connected to a photodetector. The sample was placed in intermittent contact with the tip of an atomic force microscope (AFM). The confined electromagnetic field in the structure was locally perturbed and scattered by the tip. The transmission losses due to light scattering were measured by a photodetector and sent to a lock-in amplifier, which filtered the frequency of the tip oscillation. A piezoelectric stage was used to scan the confined electromagnetic field in the $y - z$ plane.

### 3. Photoluminescence

The nanocrystals placed on top of the HS [Fig. 1(b)] were excited with a 532 nm laser to the input of the waveguide. We used a homemade confocal system with an 80 × microscope objective coupled to a 25 cm focal length spectrometer (Princeton Instruments) attached to a CDD camera cooled at −75°C. The CCD camera allowed us to image the surface and measure the spectrum of the nanocrystals from a single spot of the image. This standard microphotoluminescence system is sensitive enough to carry on lifetime measurements.

### 4. Fluorescence Lifetime

Measurements of the fluorescence lifetime of the nanoemitters were conducted by an external excitation of the nanoemitters. This excitation was made with a pulsed laser centered at $\lambda$ 405 nm with a pulse width of FWHM 55 ps, a repetition rate of 8 MHz, and an average power of around 40 μW, which was focused on the surface of the sample through a confocal system. The fluorescence of the nanocrystals due to the pulses was measured with a photocorrelator fixed at $\lambda$ 590 nm (time resolution of 264 ps). An integration time of 1 min was used to measure the fluorescence lifetime.

### 5. Purcell Factor

Purcell factor simulations were performed with a commercial software (RSoft) based on the finite-difference time-domain method (FDTD) [54,55]. A single nanocrystal was simulated as a dipolar source placed at 15 nm from the surface of the sample (half of the average radius of the nanocrystals). The Purcell factor was obtained by measuring the light radiated by the dipole oscillating in the three coordinate axes to determine the local density of excited states LDOS.

**Funding.** Agence Nationale de la Recherche (ANR) (ANR-12-NANO-0019); European Cooperation in Science and Technology (COST) (MP1403); Labex (ANR-11-LABX-01-01).

**Acknowledgment.** We thank N. Rahbany for helping with the lifetime measurements. C. C. would like to thank A. Bramati, L. Carbone, and V. de Massimo for providing the nanocrystals. The authors thank the ANR for the Sinphonie project, the Champagne-Ardenne region for the projects NANOMAT, PLASMOBIO, and NanoGain, and acknowledge the support of COST Nanoscale Quantum Optics as well as the French Labex ACTION.

## REFERENCES

1. M. C. Wu, O. Solgaard, and J. E. Ford, "Optical mems for lightwave communication, " J. Lightwave Technol. 24, 4433‾4454 (2006).
2. R. Guo, M. Decker, F. Setzpfandt, I. Staude, D. N. Neshev, and Y. S. Kivshar, "Plasmonic fano nanoantennas for on-chip separation of wavelength-encoded optical signals," Nano Lett. 15, 3324‾3328 (2015).
3. C. Zhao and J. Zhang, "Plasmonic demultiplexer and guiding," ACS Nano 4, 6433‾6438 (2010).
4. D. Pile, "Integrated photonics: compact multiplexing," Nat. Photonics 9, 78 (2015).
5. R. Stabile, A. Albores-Mejia, A. Rohit, and K. A. Williams, "Integrated optical switch matrices for packet data networks," Microsyst. Nanoeng. 2, 15042 (2016).
6. D. L. Sounas and A. Alù, "Angular-momentum-biased nanorings to realize magnetic-free integrated optical isolation, " ACS Photon. 1, 198‾204 (2014).
7. C. Sayrin, C. Junge, R. Mitsch, B. Albrecht, D. O'shea, P. Schneeweiss, J. Volz, and A. Rauschenbeutel, "Nanophotonic optical isolator controlled by the internal state of cold atoms," Phys. Rev. X 5, 041036 (2015).
8. G. T. Reed, G. Mashanovich, F. Y. Gardes, and D. J. Thomson, "Silicon optical modulators," Nat. Photonics 4, 518‾526 (2010).
9. R. J. A. Liu, L. Liao, D. Samara-Rubio, D. Rubin, O. Cohen, R. Nicolaescu, and M. Paniccia, "A high-speed silicon optical modulator based on a metal-oxide-semiconductor capacitor," Nature 427, 615‾618 (2004).
10. F. Prieto, B. Sepúlveda, A. Calle, A. Llobera, C. Domínguez, A. Abad, A. Montoya, and L. M. Lechuga, "An integrated optical interferometric nanodevice based on silicon technology for biosensor applications," Nanotechnology 14, 907‾912 (2003).
11. A. Ymeti, J. Greve, P. V. Lambeck, T. Wink, S. W. van Hövell, T. A. Beumer, R. R. Wijn, R. G. Heideman, V. Subramaniam, and J. S. Kanger, "Fast, ultrasensitive virus detection using a young interferometer sensor, " Nano Lett. 7, 394‾397 (2007).
12. R. Bruck and R. Hainberger, "Sensitivity and design of gratingassisted bimodal interferometers for integrated optical biosensing," Opt. Express 22, 32344‾32352 (2014).
13. A. Wilk, J. C. Carter, M. Chrisp, A. M. Manuel, P. Mirkarimi, J. B. Alameda, and B. Mizaikoff, "Substrate-integrated hollow waveguides: a new level of integration in mid-infrared gas sensing," Anal. Chem. 85 , 11205‾11210 (2013).
14. A. Airoudj, D. Debarnot, B. Bêche, and F. Poncin-Epaillard,  "Design and sensing properties of an integrated optical gas sensor based on a multilayer structure," Anal. Chem. 80, 9188‾9194 (2008).
15. H. Ablat, A. Yimit, M. Mahmut, and K. Itoh, "Nafion film/K+- exchanged glass optical waveguide sensor for BTX detection," Anal. Chem. 80, 7678‾7683 (2008).
16. J. Barthes, A. Bouhelier, A. Dereux, and G. C. des Francs, "Coupling of a dipolar emitter into one-dimensional surface plasmon," Sci. Rep. 3, 2734 (2013).
17. A. G. Curto, G. Volpe, T. H. Taminiau, M. P. Kreuzer, R. Quidant, and N. F. van Hulst, "Unidirectional emission of a quantum dot coupled to a nanoantenna," Science 329, 930‾933 (2010).
18. J. Fulmes, R. Jäger, A. Bräuer, C. Schäfer, S. Jäger, D. A. Gollmer, A. Horrer, E. Nadler, T. Chassé, D. Zhang, A. J. Meixner, D. P. Kern, and M. Fleischer, "Self-aligned placement and detection of quantum dots on the tips of individual conical plasmonic nanostructures," Nanoscale 7, 14691‾14696 (2015).
19. R. Tellez-Limon, B. Bahari, L. Hsu, J.-H. Park, A. Kodigala, and B. Kanté, "Integrated metaphotonics: symmetries and confined excitation of LSP resonances in a single metallic nanoparticle," Opt. Express 24, 13875‾13880 (2016).
20. R. K. Kramer, N. Pholchai, V. J. Sorger, T. J. Yim, R. Oulton, and X. Zhan, "Positioning of quantum dots on metallic nanostructures," Nanotechnology 21, 145307 (2010).
21. M. Hiscocks, C. Su, B. C. Gilbson, A. Greentree, L. Hollenberg, and F. Ladouceurl, "Slot-waveguide cavities for optical quantum information applications," Opt. Express 17, 7295‾7303 (2009).
22. P. E. Barclay, K.-M. Fu, C. Santori, and R. G. Beausoleil, "Hybrid photonic crystal cavity and waveguide for coupling to diamond NVcenters," Opt. Express 17, 9588‾9601 (2009).
23. K. Hennessy, A. Badolato, M. Winger, D. Gerace, M. Atatüre, S. Gulde, S. Fält, E. L. Hu, and A. Imamoglu, "Quantum nature of a strongly coupled single quantum dot-cavity system," Nature 445, 896‾899 (2007).
24. C. Ropp, Z. Cummins, S. Nah, J. T. Fourkas, B. Shapiro, and E. Waks, "Nanoscale imaging and spontaneous emission control with a single nano-positioned quantum dot," Nat. Commun. 4, 1447 (2013).
25. J. de Torres, P. Ferrand, G. C. des Francs, and J. Wenger, "Coupling emitters and silver nanowires to achieve long-range plasmonmediated fluorescence energy transfer," ACS Nano 10, 3968‾3976 (2016).
26. W. Geng, M. Manceau, N. Rahbany, V. Sallet, M. De Vittorio, L. Carbone, Q. Glorieux, A. Bramati, and C. Couteau, "Localized excitation of a single photon source by a nanowaveguide," Sci. Rep. 6, 19721 (2016).
27. E. Bermudez-Ureña, C. Gonzalez-Ballestero, M. Geiselmann, R. Marty, I. P. Radko, T. Holmgaard, Y. Alaverdyan, E. Moreno, F. J. García-Vidal, S. I. Bozhevolnyi, and R. Quidant, "Coupling of individual quantum emitters to channel plasmons," Nat. Commun. 6, 7883 (2015).
28. S. J. P. Kress, F. V. Antolinez, P. Richner, S. V. Jayanti, D. K. Kim, F. Prins, A. Riedinger, M. P. C. Fischer, S. Meyer, K. M. McPeak, D. Poulikakos, and D. J. Norris, "Wedge waveguides and resonators for quantum plasmonics," Nano Lett. 15, 6267‾6275 (2015).
29. Z. Fang, S. Huang, F. Lin, and X. Zhu, "Color-tuning and switching optical transport through CdS hybrid plasmonic waveguide," Opt. Express 17, 20327‾20332 (2009).
30. A. Tervonen, B. R. West, and S. Honkanen, "Ion-exchanged glass waveguide technology: a review," Opt. Eng. 50, 071107 (2011).
31. G. Zhang, S. Honkanen, A. Tervonen, C.-M. Wu, and S. I. Najafi, "Glass integrated optics circuit for 1.48/1.55- and 1.30/1.55micrometers-wavelength division multiplexing and 1/8 splitting," Appl. Opt. 33, 3371‾3374 (1994).
32. B. Pantchev and Z. Nikolov, "Characterization of refractive index profiles in silver-sodium ion-exchanged glass waveguides for homogeneous refracting waveguide structures," IEEE J. Quantum Electron. 29, 2459‾2465 (1993).

33. Z. Qi, I. Honma, and H. Zhou, "Tin-diffused glass slab waveguides locally covered with tapered thin $TiO_2$ films for application as a polarimetric interference sensor with an improved performance," Anal. Chem. 77, 1163–1166 (2005).
34. R. Rogoziński, "Ion exchange in glass—the changes of glass refraction," in Ion Exchange Technologies, A. Kilislioglu, ed., 1st ed. (InTech, 2012).
35. S. I. Najafi, P. Lefebvre, J. Albert, S. Honkanen, A. Vahid-Shahidi, and W.-J. Wang, "Ion-exchanged Mach–Zehnder interferometers in glass," Appl. Opt. 31, 3381–3383 (1992).
36. Z. Qi, S. Zhao, F. Chen, R. Liu, and S. Xia, "Performance investigation of an integrated young interferometer sensor using a novel prismchamber assembly," Opt. Express 18, 7421–7426 (2010).
37. D. Barbier, M. Rattay, F. S. Andre, G. Clauss, M. Trouillon, A. Kevorkian, J.-M. P. Delavaux, and E. Murphy, "Amplifying fourwavelength combiner, based on erbium/ytterbium-doped waveguide amplifiers and integrated splitters," IEEE Photon. Technol. Lett. 9, 315–317 (1997).
38. S. Blaize, L. Bastard, C. Cassagnetes, and J. E. Broquin, "Multiwavelengths DFB waveguide laser arrays in Yb-Er codoped phosphate glass substrate," IEEE Photon. Technol. Lett. 15, 516–518 (2003).
39. F. Gardillou and J.-E. Broquin, "Optical amplifier made by reporting an $Er^{3+}/Yb^{3+}$-codoped glass layer on an ion-exchanged passive glass substrate by wafer bonding," Proc. SPIE 5728, 120–128 (2005).
40. J. Dostalek, J. Ctyroky, J. Homola, E. Brynda, M. Skalsky, P. Nekvindova, J. Spirkova, J. Skvor, and J. Schrofel, "Surface plasmon resonance biosensor based on integrated optical waveguide," Sens. Actuators B 76, 8–12 (2001).
41. "Teem Photonics, ion-exchanged technology," http://www.teemphotonics.com/ioc/offer-and-products/ionex-technology.html, accessed 20 June 2016.
42. A. Quaranta, A. Rahman, G. Mariotto, C. Maurizio, E. Trave, F. Gonella, E. Cattaruzza, E. Gibaudo, and J. E. Broquin, "Spectroscopic investigation of structural rearrangements in silver ion-exchanged silicate glasses," J. Phys. Chem. C 116, 3757–3764 (2012).
43. M. Z. Alam, J. Meier, J. S. Aitchison, and M. Mojahedi, "Propagation characteristics of hybrid modes supported by metal-low-high index waveguides and bends," Opt. Express 18, 12971–12979 (2010).
44. R. Halir, P. J. Bock, P. Cheben, A. Ortega-Moñux, C. Alonso-Ramos, J. H. Schmid, J. Lapointe, D.-X. Xu, J. G. Wangüemert-Pérez, I. N. Molina-Fernández, and S. Janz, "Waveguide sub-wavelength structures: a review of principles and applications," Laser Photon. Rev. 9, 25–49 (2015).
45. A. A. Salman, A. Tortschanoff, G. van der Zwan, F. van Mourik, and M. Chergui, "A model for the multi-exponential excited-state decay of CdSe nanocrystals," Chem. Phys. 357, 96–101 (2009).
46. G. Sagarzazu, K. Inoue, M. Saruyama, M. Sakamoto, T. Teranishi, S. Masuo, and N. Tamai, "Ultrafast dynamics and single particle spectroscopy of Au–CdSe nanorods," Phys. Chem. Chem. Phys. 15, 2141–2152 (2013).
47. R. Carminati, A. Cazé, D. Cao, F. Peragut, V. Krachmalnicoff, R. Pierrat, and Y. D. Wilde, "Electromagnetic density of states in complex plasmonic systems," Surf. Sci. Rep. 70, 1–41 (2015).
48. N. Chateau and J. P. Hugonin, "Algorithm for the rigorous coupledwave analysis of grating diffraction," J. Opt. Soc. Am. A 11, 1321–1331 (1994).
49. L. Li, "Formulation and comparison of two recursive matrix algorithms for modeling layered diffraction gratings," J. Opt. Soc. Am. A 13, 1024–1035 (1996).
50. P. Lalanne and E. Silberstein, "Fourier-modal methods applied to waveguide computational problems," Opt. Express 25, 1092–1094 (2000).
51. R. Tellez-Limon, M. Fevrier, A. Apuzzo, R. Salas-Montiel, and S. Blaize, "Theoretical analysis of Bloch mode propagation in an integrated chain of gold nanowires," Photon. Res. 2, 24–30 (2014).
52. L. Gomez, R. Bachelot, A. Bouhelier, G. P. Wiederrecht, S. Chang, S. K. Gray, F. Hua, S. Jeon, J. A. Rogers, M. E. Castro, S. Blaize, I. Stefanon, G. Lerondel, and P. Royer, "Apertureless scanning near-field optical microscopy: a comparison between homodyne and heterodyne approaches," J. Opt. Soc. Am. B 23, 823–833 (2006).
53. S. Blaize, S. Aubert, A. B. R. Bachelot, G. Lerondel, P. Royer, J.-E. Broquin, and V. Minier, "Apertureless scanning near-field optical microscopy for ion exchange channel waveguide characterization," J. Microsc. 209, 155–161 (2003).
54. A. Taflove and S. Hagness, Computational Electrodynamics: The Finite-Difference Time-Domain Method (Artech House, 2000).
55. P. T. Kristensen and S. Hughes, "Modes and mode volumes of leaky optical cavities and plasmonic nanoresonators," ACS Photon. 1, 2–10 (2014).